\pdfoutput=1

\documentclass[preprint,showpacs,preprintnumbers,amsmath,amssymb,floatfix,endfloats*]{revtex4}

\usepackage{graphicx}
\usepackage{dcolumn}
\usepackage{bm}
\usepackage{amsfonts}

\DeclareMathAlphabet{\mathsfsl}{OT1}{cmr}{bx}{it}
\begin{document}
\title{Structural transformations in porous glasses under mechanical loading. I. Tension}
\author{Nikolai V. Priezjev$^{1,2}$ and Maxim A. Makeev$^{3}$}
\affiliation{$^{1}$Department of Mechanical and Materials
Engineering, Wright State University, Dayton, OH 45435}
\affiliation{$^{2}$National Research University Higher School of
Economics, Moscow 101000, Russia}
\affiliation{$^{3}$Department of Chemistry, University of
Missouri-Columbia, Columbia, MO 65211}
\date{\today}
\begin{abstract}

The evolution of porous structure and mechanical properties of
binary glasses under tensile loading were examined using molecular
dynamics simulations.   We consider vitreous systems obtained in the
process of phase separation after a rapid isochoric quench of a
glass-forming liquid to a temperature below the glass transition.
The porous structure in undeformed samples varies from a connected
porous network to a random distribution of isolated pores upon
increasing average glass density.  We find that at small strain, the
elastic modulus follows a power-law dependence on the average glass
density and the pore size distribution remains nearly the same as in
quiescent samples.  Upon further loading, the pores become
significantly deformed and coalesce into larger voids that leads to
formation of system-spanning empty regions associated with breaking
of the material.

\end{abstract}



\maketitle

\section{Introduction}

Recent progress in the development of porous structural materials
with applications ranging from biomedicine to energy conversion and
storage as well as civil infrastructure, requires a thorough
understanding of their microstructure--property
relationship~\cite{McKittrick14,Tomsia11,LiuBio13,Okoli14,Su12}.  An
accurate pore characterization of microporous materials, which
involves numerical evaluation of the probe-accessible and
-occupiable pore volume, allows determination of their permeability
to guest molecules and internal void volume and surface
area~\cite{Haranczyk17}.  The results of experimental and
computational studies have shown that mechanical properties of bulk
metallic glasses with periodic arrays of pores are governed by shear
localization between adjacent pores in a regime of plastic
deformation~\cite{Bargmann14,EckertMet16}. Similar to ductile
metallic alloys, it was found that in highly strained nanoporous
silica glasses, multiple cracks are initiated at void surfaces,
which leads to void coalescence and intervoid ligament
failure~\cite{Vashishta07}.   It was further shown that mechanical
properties of porous silica glasses are improved in samples with
channel pore morphology rather than isolated pore
configurations~\cite{Rimsza14}.   Despite extensive efforts, the
precise connection between pore morphology and elastic, shear and
bulk moduli has not yet been determined.

\vskip 0.05in

During the last decade, the mechanical properties of metallic glass
nanowires subjected to uniaxial tension have been investigated using
molecular dynamics simulations~\cite{Shi10,Yuan12,Eckert16,MoLi17}
and experimental
measurements~\cite{MaNature07,Greer13,Gianola13,Lewand15}. It was
observed that when the size of metallic glass samples is reduced
down to the nanoscale, the deformation mode changes from brittle to
ductile~\cite{Eckert16,MoLi17,MaNature07,Greer13,Gianola13,Lewand15}.
The difference in the deformation behavior can be visually detected
by observing either shear localization along a plane, called a shear
band, or the formation of extended necking along the loading
direction~\cite{Shi10,Greer13}.    A subsequent analysis of
irradiated samples that were emulated in MD simulations by randomly
removing a small fraction of atoms, has shown an enhanced tensile
ductility; while this effect is reduced if only the outer shell of a
nanowire is rejuvenated~\cite{Eckert16}.   It was also demonstrated
that a homogeneous bulk metallic glass under uniaxial tension
exhibits only one dominant shear band, whereas multiple shear bands
are initiated at interfaces between grains in a
nanoglass~\cite{Albe11}. Moreover, it was found that the shear-band
direction with respect to the loading direction is different in the
cases of uniaxial compression and extension of two-dimensional
athermal amorphous solids~\cite{Gendelman13}.  However, the
combination of several factors including the processing routes,
system size and aspect ratio as well as surface defects and local
microstructure makes it difficult to predict accurately the failure
mode in strained glasses.

\vskip 0.05in

A few years ago, the liquid-gas phase separation kinetics of a
glass-forming system quenched rapidly from a liquid state to a
temperature below the glass transition was studied via molecular
dynamics simulations~\cite{Kob11,Kob14}.  As a results of the
coarsening process at constant volume, an amorphous solid is formed,
whose porous structure contains isolated voids at higher average
glass densities and complex interconnected morphologies at lower
glass densities~\cite{Kob11,Kob14}.   More recently, the
distributions of pore sizes and local glass densities were further
investigated as a function of temperature and average glass
density~\cite{Makeev17}. In particular, it was found that in systems
with high porosity, the pore size distribution functions obey a
scaling relation up to intermediate length scales, while in highly
dense systems, the distribution is nearly Gaussian~\cite{Makeev17}.
Furthermore, under steady shear deformation, the pores become
significantly deformed and, at large strain, they were shown to
aggregate into large voids that are comparable with the system
size~\cite{Foffi14,Priezjev17}. It was also demonstrated that the
shear modulus follows a power-law dependence as a function of the
average glass density~\cite{Priezjev17}.  Nevertheless, the
mechanical response of porous glasses to different types of loading
conditions and the transformation of porous structure remain not
fully understood.

\vskip 0.05in

In this paper, molecular dynamics simulations are carried out to
investigate the pore size distribution and mechanical properties of
a model glass under tensile deformation. The porous glass is
prepared via a deep quench of a binary mixture in a liquid state to
a very low temperature at constant volume. It will be shown that
under tension, the distribution of pore sizes becomes highly skewed
towards larger values, and upon further increasing strain, one large
dominant pore is formed in the region where failure occurs. The
analysis of local density profiles and visualization of atom
configurations reveals that the location of the failure zone is
correlated with the extent of a lower glass density region.

\vskip 0.05in

The rest of the paper is structured as follows. In the next section,
we describe the details of molecular dynamics simulations including
model parameters as well as equilibration and deformation protocols.
The results for the stress-strain response, evolution of density
profiles and pore size distributions are presented in
Sec.\,\ref{sec:Results}. A brief summary and outlook are given in
the last section.

\section{Details of molecular dynamics simulations}
\label{sec:MD_Model}


The mechanical properties of porous glasses were investigated using
the standard Kob-Andersen (KA) binary (80:20) mixture
model~\cite{KobAnd95}.  In this model, the interaction between any
two atoms are described via the Lennard-Jones (LJ) potential:
\begin{equation}
V_{\alpha\beta}(r)=4\,\varepsilon_{\alpha\beta}\,\Big[\Big(\frac{\sigma_{\alpha\beta}}{r}\Big)^{12}\!-
\Big(\frac{\sigma_{\alpha\beta}}{r}\Big)^{6}\,\Big],
\label{Eq:LJ_KA}
\end{equation}
where the parameters are set to $\varepsilon_{AA}=1.0$,
$\varepsilon_{AB}=1.5$, $\varepsilon_{BB}=0.5$, $\sigma_{AB}=0.8$,
$\sigma_{BB}=0.88$, and $m_{A}=m_{B}$~\cite{KobAnd95}. For
computational efficiency, the LJ forces were only computed at
distances smaller than the cutoff radius
$r_{c,\,\alpha\beta}=2.5\,\sigma_{\alpha\beta}$. In what follows,
the LJ units of length, mass, energy, and time are
$\sigma=\sigma_{AA}$, $m=m_{A}$, $\varepsilon=\varepsilon_{AA}$,
and, consequently,  $\tau=\sigma\sqrt{m/\varepsilon}$. The equations
of motion were solved numerically using the Verlet
algorithm~\cite{Allen87}, implemented in LAMMPS~\cite{Lammps}, with
the time step $\triangle t_{MD}=0.005\,\tau$.

\vskip 0.05in


Our model porous systems were prepared by first equilibrating
$N=300\,000$ atoms in a cubic cell at the temperature of
$1.5\,\varepsilon/k_B$ during $3\times 10^4\,\tau$, while keeping
the volume constant. Here, $k_B$ denotes the Boltzmann constant. For
reference, the computer glass transition temperature of the KA model
is $T_g\approx0.435\,\varepsilon/k_B$~\cite{KobAnd95}.  Second, the
temperature of the liquid phase was instantaneously set to the low
value of $0.05\,\varepsilon /k_{B}$, and the system was evolving
during the time interval $10^{4}\,\tau$ at constant volume. During
this process, the temperature of $0.05\,\varepsilon /k_{B}$ was
maintained by simple velocity rescaling.  Examples of the resulting
porous structures are presented in Fig.\,\ref{fig:snapshot_system}
for the average glass densities $\rho\sigma^{3}=0.2$, $0.4$, $0.6$
and $0.8$.   The equilibration and quenching procedures were
performed for the average glass densities in the range
$0.2\leq\rho\sigma^{3} \leq 1.0$, and the data were averaged in five
independent samples for each value of $\rho\sigma^{3}$.   Note that
sample preparation protocols are the same as in the previous MD
studies~\cite{Makeev17,Priezjev17}.

\vskip 0.05in


In the next step, the porous samples were strained along the
$\hat{x}$ direction with the strain rate
$\dot{\varepsilon}_{xx}=10^{-4}\,\tau^{-1}$ while being compressed
along the $\hat{y}$ and $\hat{z}$ directions in order to keep the
volume constant.  The deformation takes place during the time
interval of $10^{4}\,\tau$ at the temperature of $0.05\,\varepsilon
/k_{B}$, which is regulated by the Nos\'{e}-Hoover
thermostat~\cite{Lammps}. In nonequilibrium simulations, all stress
components and system dimensions were saved every $0.5\,\tau$ as
well as atomic configurations that were stored every $500\,\tau$.
These data were analyzed in five independent samples for each value
of the average glass density, and the results for the pore size
distribution, density profiles, and the elastic modulus are
presented in the next section.

\section{Results}
\label{sec:Results}

\begin{figure}[t]
\includegraphics[width=15.cm,angle=0]{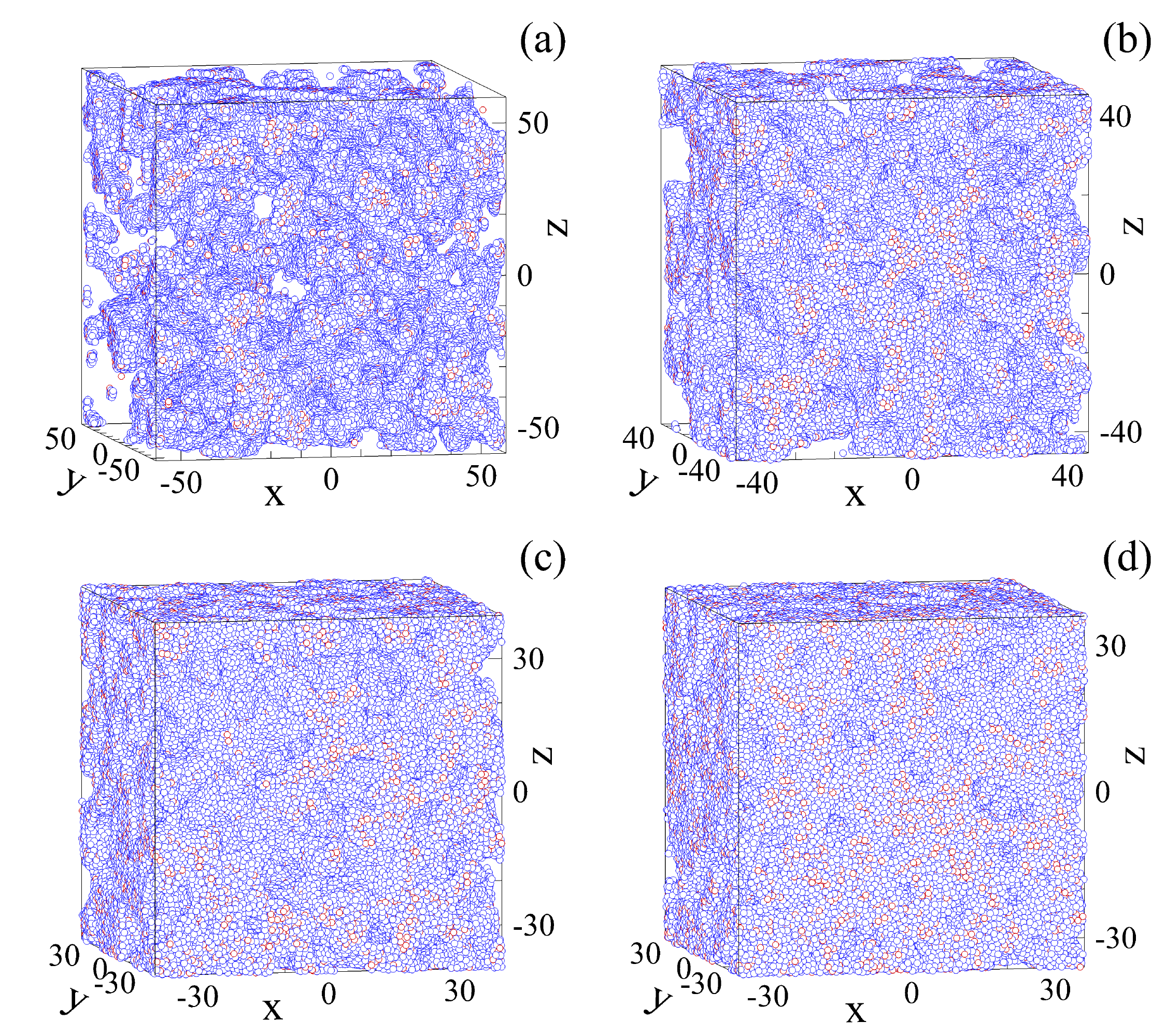}
\caption{(Color online) Atom positions in the porous binary glass
after isochoric quench to the temperature $T=0.05\,\varepsilon/k_B$
for the average glass densities (a) $\rho\sigma^{3}=0.2$, (b)
$\rho\sigma^{3}=0.4$, (c) $\rho\sigma^{3}=0.6$, and (d)
$\rho\sigma^{3}=0.8$. The blue and red circles indicate atom types A
and B. The total number of atoms is $N=300\,000$. Note that atoms
are not drawn to scale and the system sizes are different in all
panels.}
\label{fig:snapshot_system}
\end{figure}

It was recently demonstrated that when a glass-forming system is
rapidly quenched at constant volume from a high-temperature liquid
state to a temperature below the glass transition, the porous
structures are developed at sufficiently low average glass densities
and fast cooling rates~\cite{Kob11,Kob14}.   In the previous MD
studies, the distribution of pore sizes and local densities of the
solid phase~\cite{Makeev17} as well as temporal evolution of pore
sizes and mechanical properties of systems under steady
shear~\cite{Priezjev17} were investigated in a wide range of average
glass densities $0.2 \leqslant \rho\sigma^{3} \leqslant 1.0$.
Following the preparation procedure described in
Sec.\,\ref{sec:MD_Model}, we show the representative snapshots of
the porous samples before deformation in
Fig.\,\ref{fig:snapshot_system} for the average glass densities
$\rho\sigma^{3}=0.2$, $0.4$, $0.6$ and $0.8$.  It can be observed
that at higher glass densities $\rho\sigma^{3} \gtrsim 0.6$, the
porous structure involves isolated voids with various sizes up to
several molecular diameters, while at lower densities, the average
size of pores increases and they become more interconnected, with
channels running through the systems. Notice that at the lower glass
density $\rho\sigma^{3}=0.2$ in
Fig.\,\ref{fig:snapshot_system}\,(a), there are a number of straight
paths across the whole sample, indicating that the porous network is
above the percolation threshold.

\vskip 0.05in


The formation of porous glass occurs at a constant volume, and,
therefore, the systems undergo evolution under the condition of
negative pressure~\cite{Paluch15,Paluch17}. This has a number of
important implications for the thermodynamic states of the porous
glassy systems. First, the systems under consideration exist in
metastable states. Also note that these systems can be envisaged as
effectively confined~\cite{Paluch15,Paluch17}. Therefore, there
exist a distribution of built-in tensile stresses in the solid
domains of each binodal system. Correspondingly, the effects due to
these stress distributions are expected to contribute to the
dynamical evolution of the systems under mechanical loading; i.e.,
when the thermodynamic barriers are perturbed by an applied external
load. As was discussed in Ref.\,\cite{Makeev17}, the spinodal
decomposition during the transition from liquid phase to that of
porous glass occurs such that an extended high density domains are
formed in the systems. In the process of elongation (tension), a
farther phase separation can be made possible due to lowering of the
corresponding thermodynamic barriers. Thus, the expected behavior is
a redistribution of material from the region close to the failure
zone to remote domains.   In some cases, this process can be
accompanied by pore shrinkage and/or closure. That is what we
observed in the present study, as is detailed below. In turn, these
structural transformations should lead to a significant decrease in
built-in stresses.

\vskip 0.05in

\begin{figure}[t]
\includegraphics[width=12.cm,angle=0]{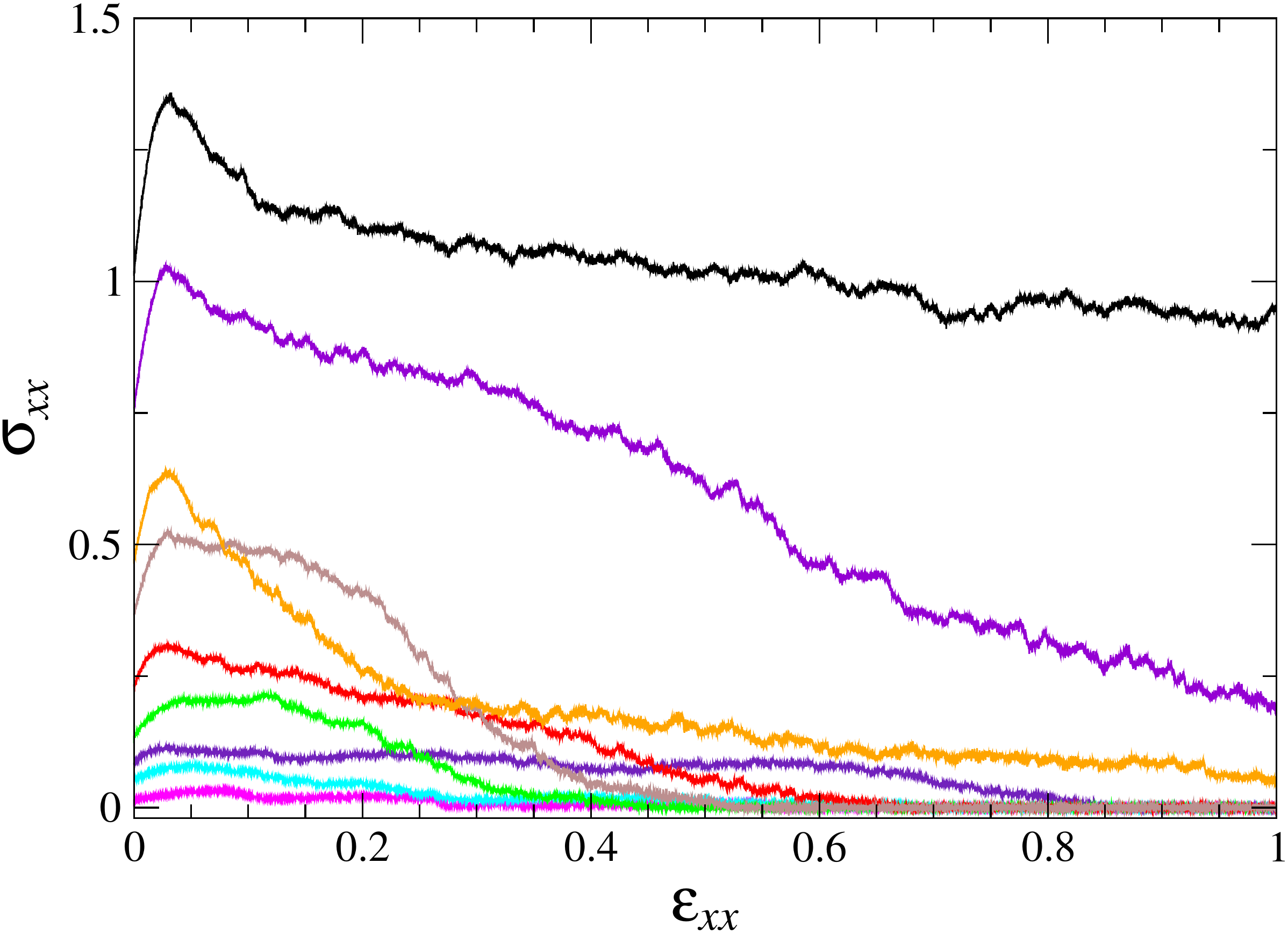}
\caption{(Color online) The dependence of stress $\sigma_{xx}$ (in
units of $\varepsilon\sigma^{-3}$) as a function of strain for the
average glass densities $\rho\sigma^{3}=0.2$, $0.3$, $0.4$, $0.5$,
$0.6$, $0.7$, $0.8$, $0.9$ and $1.0$ (from bottom to top). The
strain rate is $\dot{\varepsilon}_{xx}=10^{-4}\,\tau^{-1}$ and
temperature is $T=0.05\,\varepsilon/k_B$. }
\label{fig:stress_strain}
\end{figure}

The typical stress-strain curves are plotted in
Fig.\,\ref{fig:stress_strain} for $\varepsilon_{xx} \leqslant 1.0$
and $0.2 \leqslant \rho\sigma^{3} \leqslant 1.0$.   The data are
extracted from one sample for each value of the average glass
density $\rho\sigma^3$.   In our study, the tensile deformation
along the $\hat{x}$ direction was performed at a computationally
slow strain rate $\dot{\varepsilon}_{xx}=10^{-4}\,\tau^{-1}$ while
keeping the volume constant. Thus, at the end of the deformation
process, the original cell size in the $\hat{x}$ direction, $L_x$,
increases by a factor of two when $\varepsilon_{xx}=1.0$.   As shown
in Fig.\,\ref{fig:stress_strain}, the stress, $\sigma_{xx}$, at zero
strain is finite, and its magnitude increases at higher densities.
This behavior is consistent with the results of previous MD study,
where the effects of temperature and average glass density on
negative pressure in porous systems at equilibrium were thoroughly
investigated~\cite{Makeev17}. In particular, it was shown that
pressure is is a strong function of the average density at low
temperatures, and the data for different densities are well
described by the scaling relation
$P/T\sim\rho^{\alpha}$~\cite{Makeev17}.

\vskip 0.05in

\begin{figure}[t]
\includegraphics[width=12.cm,angle=0]{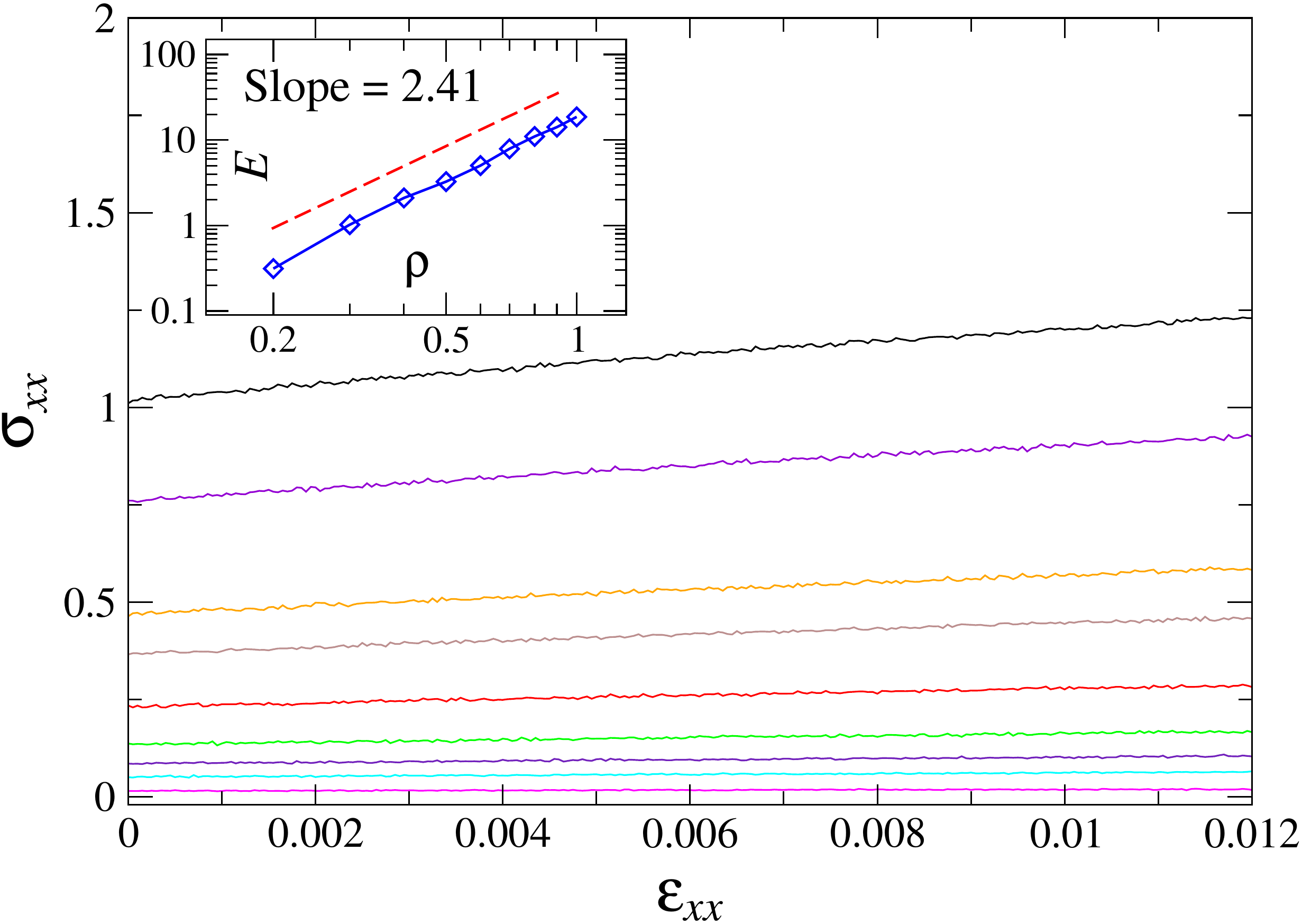}
\caption{(Color online) The enlarged view of the stress-strain
curves at strain $\varepsilon_{xx} \lesssim 0.012$ and
$\rho\sigma^{3}=0.2$, $0.3$, $0.4$, $0.5$, $0.6$, $0.7$, $0.8$,
$0.9$ and $1.0$ (from bottom to top).  The same data as in
Fig.\,\ref{fig:stress_strain}. The inset shows the elastic modulus
$E$ (in units of $\varepsilon\sigma^{-3}$) versus the average glass
density $\rho\sigma^{-3}$.   The straight dashed line denotes the
slope of $2.41$. }
\label{fig:stress_strain_1pr}
\end{figure}

At the initial stage of tensile deformation, $\varepsilon_{xx}
\lesssim 0.04$, the stress increases for each value of the average
glass density until it acquires a distinct maximum at the yield
strain (see Fig.\,\ref{fig:stress_strain}). Upon further increasing
strain, $\varepsilon_{xx} \gtrsim 0.04$, the stress gradually
decreases down to zero in porous systems with smaller average
density, indicating material's failure and breaking up into separate
domains (discussed below).  An enlarged view of the stress-strain
curves at small strain is shown in
Fig.\,\ref{fig:stress_strain_1pr}.  It can be seen that the tensile
stress, $\sigma_{xx}$, is a linear function of strain for
$\varepsilon_{xx} \lesssim 0.01$, and the slope, or the elastic
modulus, increases at higher glass densities. In the inset to
Fig.\,\ref{fig:stress_strain_1pr}, the elastic modulus, $E$,
averaged over five independent samples at each $\rho\sigma^3$, is
plotted as a function of the average glass density. The results of
our study show that the data are well described by the power-law
function, $E\sim\rho^{2.41}$ (see the red line in the inset to
Fig.\,\ref{fig:stress_strain_1pr}).  It should be emphasized that
the same exponent of $2.41$ was reported for the dependence of the
shear modulus versus glass density of porous samples under steady
shear~\cite{Priezjev17}.   We also comment that the slopes of the
stress-strain curves shown in Fig.\,\ref{fig:stress_strain_1pr}
remain the same at small negative values of strain during
compression deformation (not shown).

\vskip 0.05in

\begin{figure}[t]
\includegraphics[width=15.cm,angle=0]{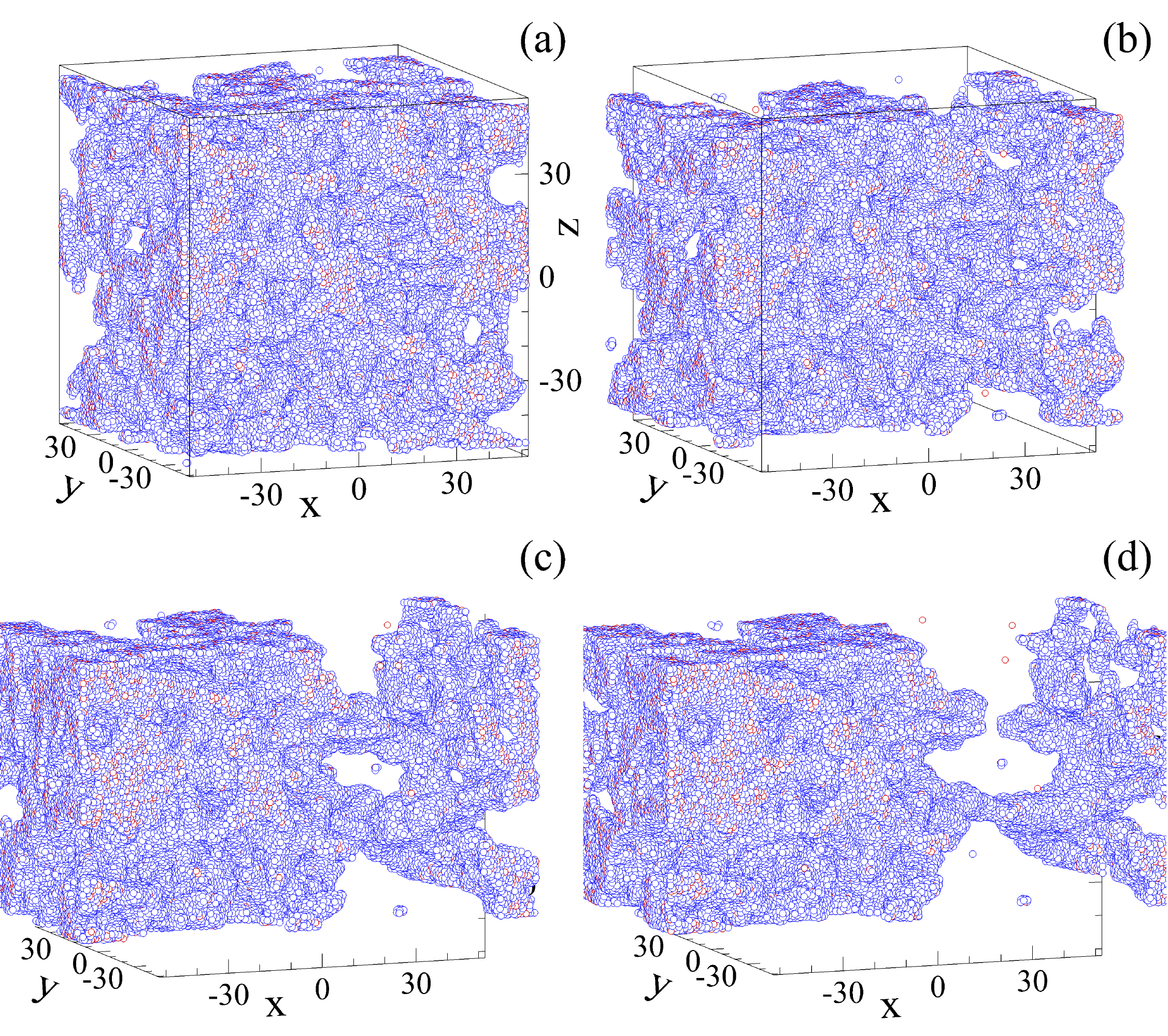}
\caption{(Color online) Instantaneous atom positions for the average
glass density $\rho\sigma^{3}=0.3$ and strain (a)
$\varepsilon_{xx}=0.05$, (b) $\varepsilon_{xx}=0.25$, (c)
$\varepsilon_{xx}=0.45$, and (d) $\varepsilon_{xx}=0.60$. The strain
rate is $\dot{\varepsilon}_{xx}=10^{-4}\,\tau^{-1}$. }
\label{fig:snapshot_strain_rho03}
\end{figure}
\begin{figure}[t]
\includegraphics[width=15.cm,angle=0]{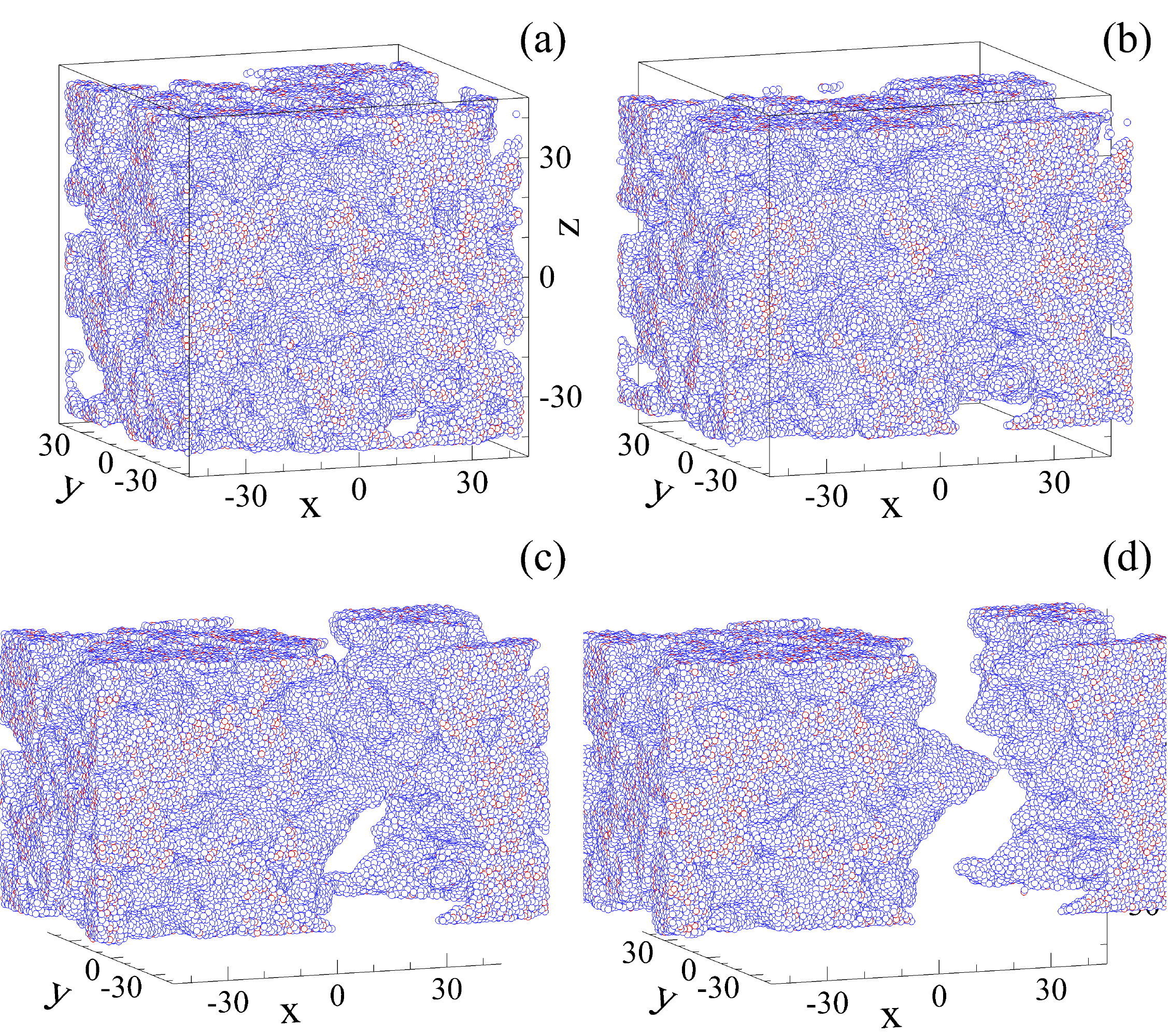}
\caption{(Color online) A sequence of atomic configurations for the
average glass density $\rho\sigma^{3}=0.5$ and strain values (a)
$\varepsilon_{xx}=0.05$, (b) $\varepsilon_{xx}=0.25$, (c)
$\varepsilon_{xx}=0.45$, and (d) $\varepsilon_{xx}=0.60$.}
\label{fig:snapshot_strain_rho05}
\end{figure}
\begin{figure}[t]
\includegraphics[width=15.cm,angle=0]{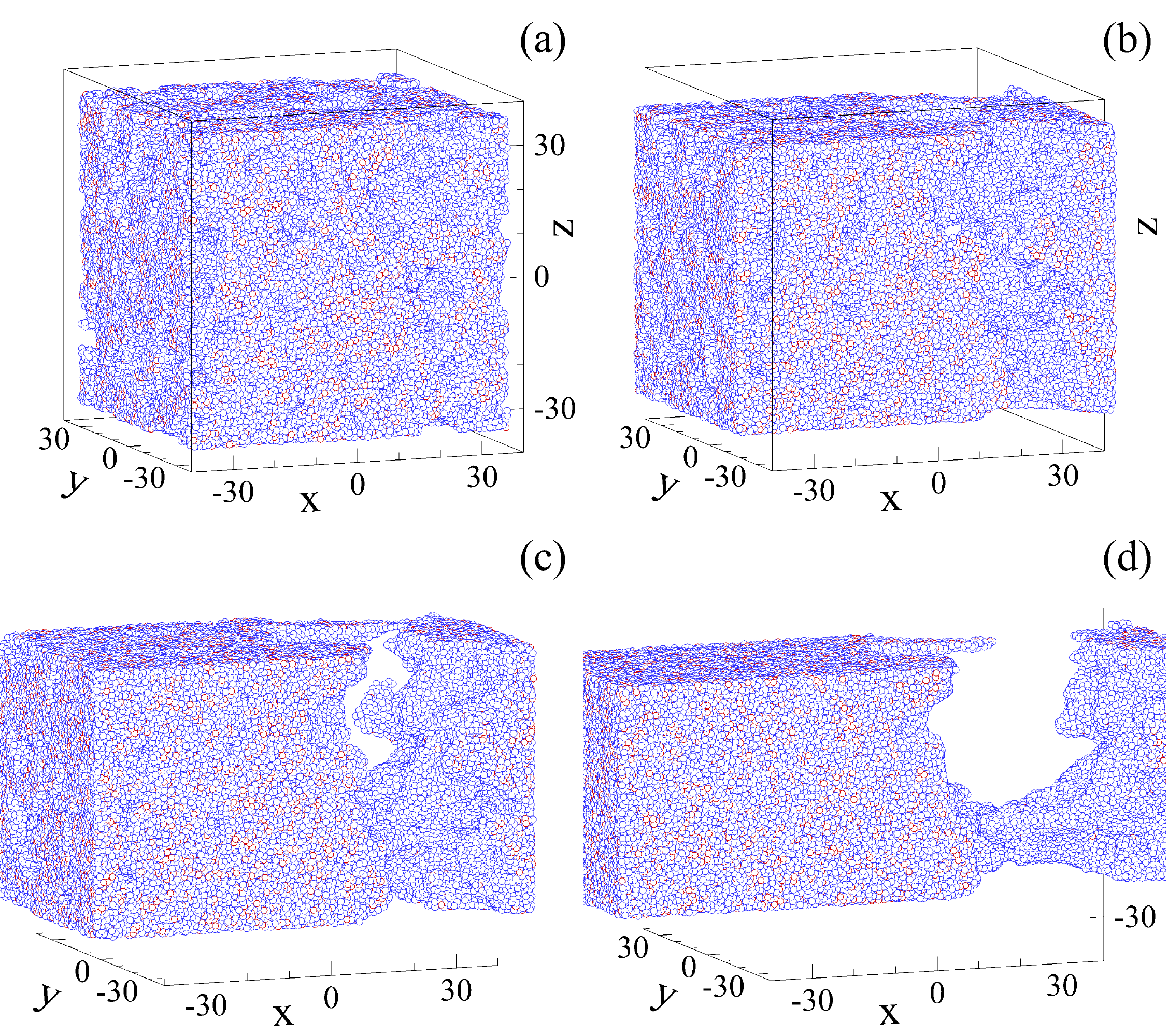}
\caption{(Color online) Snapshots of atom positions for the average
glass density $\rho\sigma^{3}=0.8$ and strain (a)
$\varepsilon_{xx}=0.05$, (b) $\varepsilon_{xx}=0.25$, (c)
$\varepsilon_{xx}=0.45$, and (d) $\varepsilon_{xx}=0.95$. }
\label{fig:snapshot_strain_rho08}
\end{figure}
\begin{figure}[t]
\includegraphics[width=15.cm,angle=0]{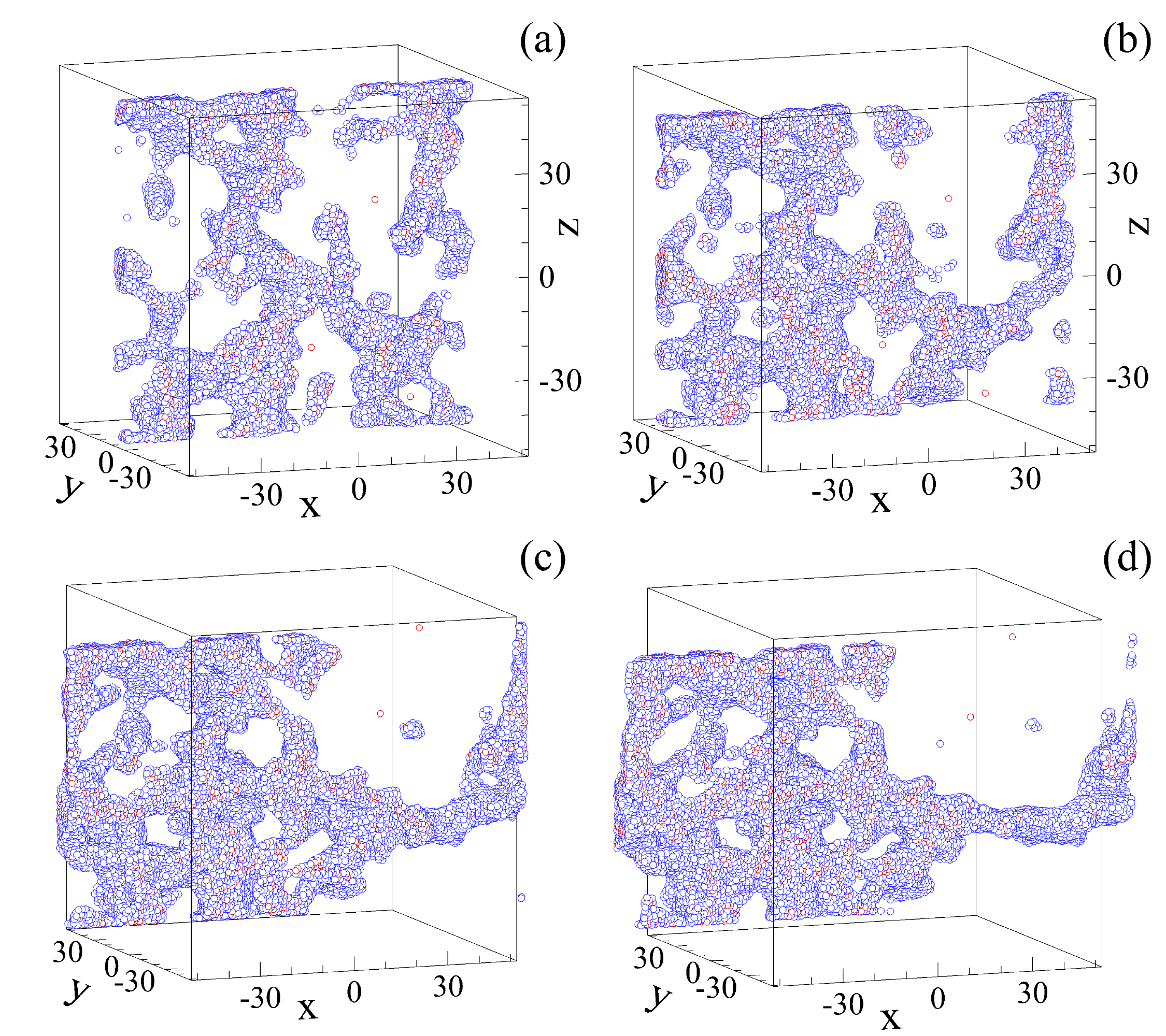}
\caption{(Color online) Atom configurations within a slice of
thickness $10\,\sigma$ for the average glass density
$\rho\sigma^{3}=0.3$ and strain (a) $\varepsilon_{xx}=0.05$, (b)
$\varepsilon_{xx}=0.25$, (c) $\varepsilon_{xx}=0.45$, and (d)
$\varepsilon_{xx}=0.60$.   Each panel contains a subset of the data
shown in Fig.\,\ref{fig:snapshot_strain_rho03}.  }
\label{fig:snapshot_strain_rho03_slice}
\end{figure}
\begin{figure}[t]
\includegraphics[width=15.cm,angle=0]{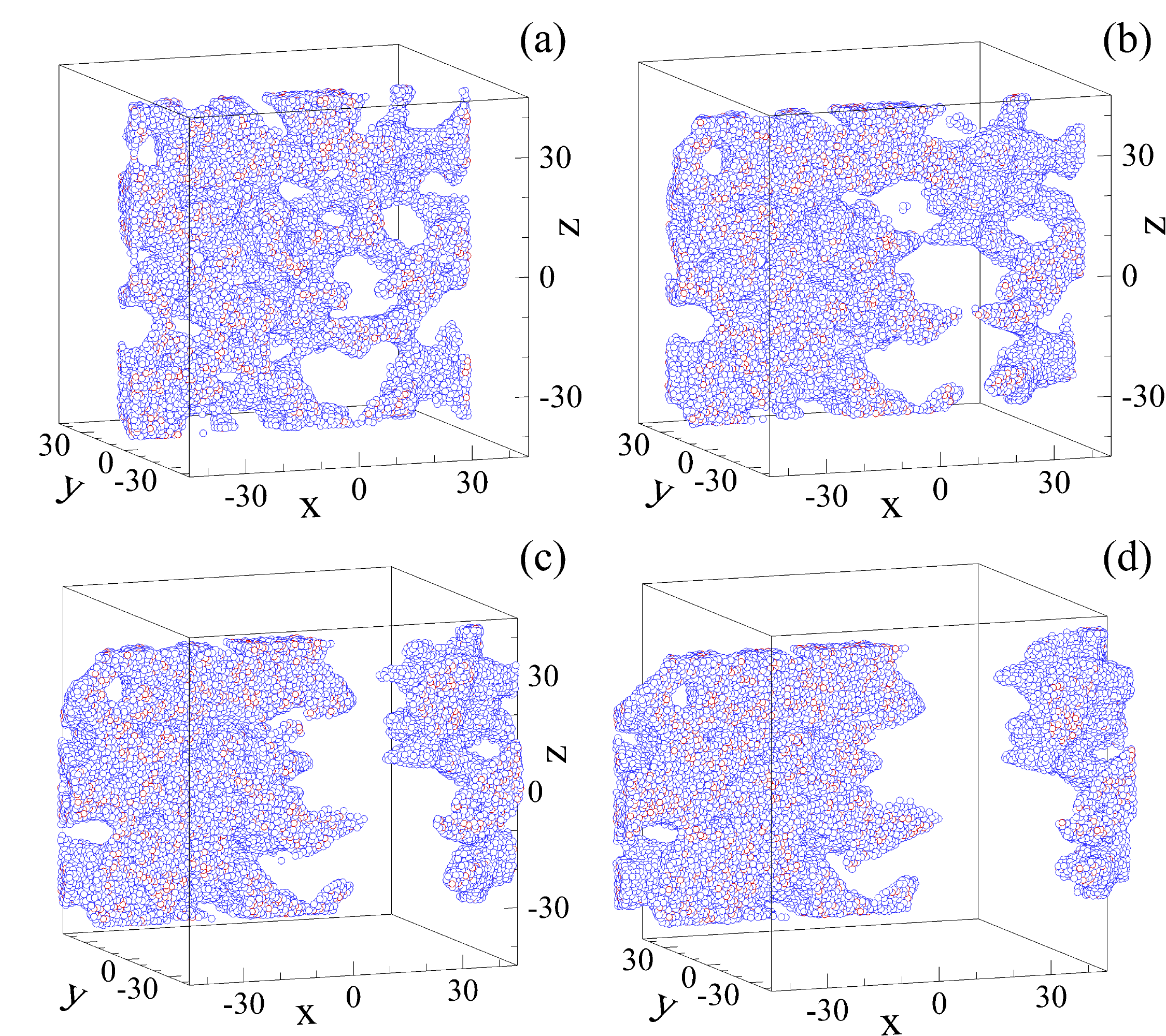}
\caption{(Color online) Atom positions within the narrow slice of
$10\,\sigma$ for the average glass density $\rho\sigma^{3}=0.5$ and
strain (a) $\varepsilon_{xx}=0.05$, (b) $\varepsilon_{xx}=0.25$, (c)
$\varepsilon_{xx}=0.45$, and (d) $\varepsilon_{xx}=0.60$. The same
data as in Fig.\,\ref{fig:snapshot_strain_rho05}. }
\label{fig:snapshot_strain_rho05_slice}
\end{figure}
\begin{figure}[t]
\includegraphics[width=15.cm,angle=0]{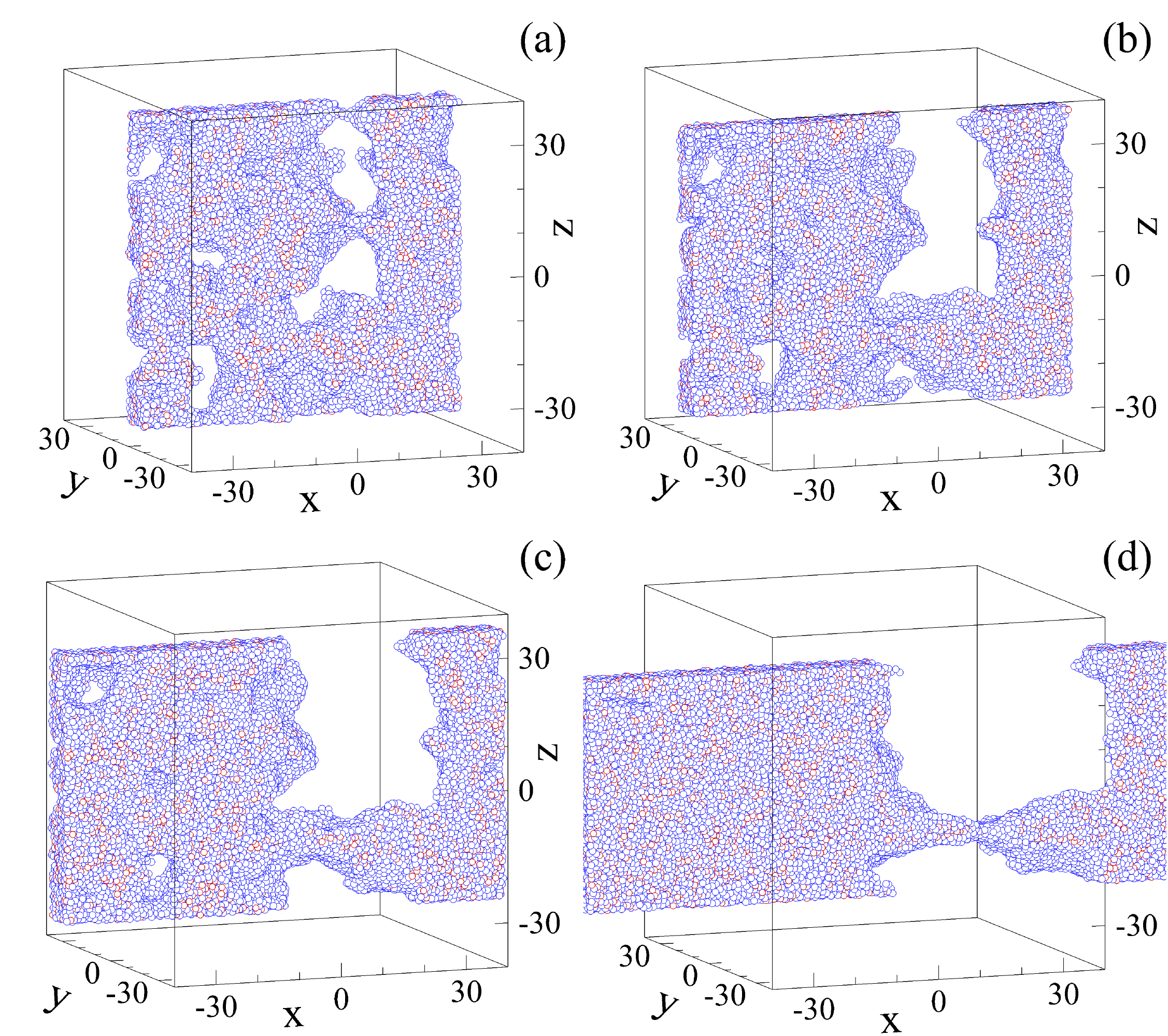}
\caption{(Color online) Snapshots of the porous glass in a thin
slice of $10\,\sigma$ for the average glass density
$\rho\sigma^{3}=0.8$ and strain (a) $\varepsilon_{xx}=0.05$, (b)
$\varepsilon_{xx}=0.25$, (c) $\varepsilon_{xx}=0.45$, and (d)
$\varepsilon_{xx}=0.95$. The same data set as in
Fig.\,\ref{fig:snapshot_strain_rho08}.   }
\label{fig:snapshot_strain_rho08_slice}
\end{figure}

The evolution of the porous structure during tensile deformation is
illustrated in Figs.\,\ref{fig:snapshot_strain_rho03},
\ref{fig:snapshot_strain_rho05} and \ref{fig:snapshot_strain_rho08}
for samples with $\rho\sigma^3=0.3$, $0.5$ and $0.8$.  In all cases,
it is evident that with increasing strain, the pore shapes become
highly distorted resulting ultimately in the formation of a single
large void that separates solid domains.   It can be observed that
in the highly strained sample with the density $\rho\sigma^3=0.8$,
shown in Fig.\,\ref{fig:snapshot_strain_rho08}\,(d), the pores are
essentially absent in the bulk glass due to compression along the
lateral dimensions. Furthermore, the finite value of tensile stress
at large strain $\varepsilon_{xx}\approx 1.0$ for $\rho\sigma^3=0.8$
in Fig.\,\ref{fig:stress_strain} is associated with formation of the
extended neck connecting solid domains shown in
Fig.\,\ref{fig:snapshot_strain_rho08}\,(d). Notice also in
Fig.\,\ref{fig:snapshot_strain_rho03} that the sample with the lower
density $\rho\sigma^3=0.3$ contains a number of small isolated
clusters of atoms in the sparse network due to the finite cutoff
radius of the LJ potential.   The transformation of pore shapes in
strained glasses can be more easily detected by visual inspection of
a sequence of atom configurations in thin slices of $10\,\sigma$
presented in Figs.\,\ref{fig:snapshot_strain_rho03_slice},
\ref{fig:snapshot_strain_rho05_slice}, and
\ref{fig:snapshot_strain_rho08_slice}. A more quantitative
description of the distribution of void space can be obtained by
counting a number of spheres of different sizes that can be inserted
into the porous structure.

\vskip 0.05in

In our numerical analysis, the pore size distribution (PSD)
functions were computed using the ZEO++
software~\cite{Haranczyk12,Haranczyk12c,Haranczyk17}. The pore sizes
were evaluated using the following computational approach. First, a
decomposition of the system volume into Voronoi cells, associated
with each individual atom, was performed. Thereby obtained Voronoi
network contains information on the nodes and edges of the Voronoi
cells. Upon successful decomposition, the total volume is equal to
the sum of the volumes of the corresponding Voronoi cells. In the
network, all nodes and edges are labeled with their distance to the
corresponding set of nearest atoms, and, thus, the obtained Voronoi
network represents void space in a porous material (that includes
isolated pores and channels). The implementation of the numerical
method is based on a variation of the Dijkstra's shortest-path
algorithm~\cite{Dijkstra59}.

\vskip 0.05in

\begin{figure}[t]
\includegraphics[width=12.cm,angle=0]{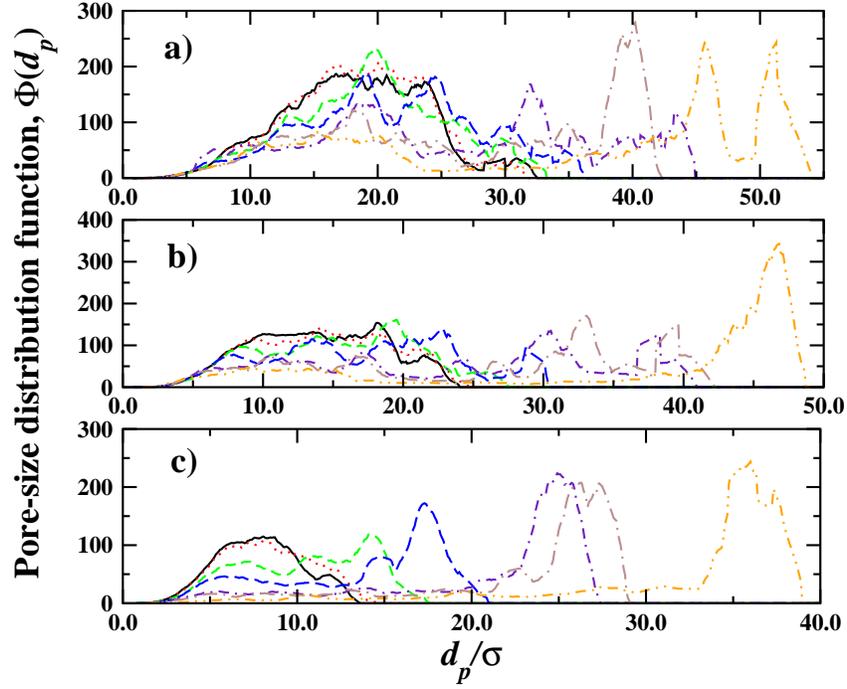}
\caption{(Color online) The distribution of pore sizes for the
average glass densities (a) $\rho\sigma^{3}=0.3$, (b)
$\rho\sigma^{3}=0.5$, and (c) $\rho\sigma^{3}=0.8$.  The
distribution functions at different strains are indicated by solid
black curves ($\varepsilon_{xx}=0.0$), dotted red curves
($\varepsilon_{xx}=0.05$), dashed green curves
($\varepsilon_{xx}=0.15$), dashed blue curves
($\varepsilon_{xx}=0.25$), dash-dotted velvet curves
($\varepsilon_{xx}=0.45$), dash-dotted brown curves
($\varepsilon_{xx}=0.50$), double-dot-dashed orange curves
($\varepsilon_{xx}=0.75$). }
\label{fig:pore_size_dist}
\end{figure}

The pore size distribution functions, $\Phi(d_p)$, are presented in
Fig.\,\ref{fig:pore_size_dist} for the average glass densities
$\rho\sigma^{3}=0.3$, $0.5$ and $0.8$.  In agreement with the
results of our previous study~\cite{Makeev17}, the distribution of
pore sizes in quiescent samples is narrow at high glass densities
and it becomes broader as the average glass density decreases. The
specifics of PSDs for porous binary glasses at equilibrium were
previously discussed by the authors in Ref.\,\cite{Makeev17}.
Subsequently, the temporal evolution of PSDs in porous glasses
undergoing steady shear were thoroughly investigated in
Ref.\,\cite{Priezjev17}. Similar to the case of simple
shear~\cite{Priezjev17}, under small strain deformation,
$\varepsilon_{xx} = 0.05$, the shape of PSD curves remains largely
unaffected (see Fig.\,\ref{fig:pore_size_dist}). With increasing
strain, the PSDs widen and start to develop a double-peak shape.  In
the regions of smaller $d_p$, the magnitude of PSDs decrease
drastically, while the magnitudes of peaks, developed at larger
values of $d_p$, increase. This type of behavior was also observed
in porous glasses under shear~\cite{Priezjev17}. In the case of
tension, however, this effect is significantly amplified in the case
of $\rho\sigma^{3}=0.8$.  Indeed, the small-size pores nearly
disappear, when $\varepsilon_{xx}$ exceeds $0.5$.  Also, in the
limit of extremely large strains ($\varepsilon_{xx} \rightarrow
1.0$), the double peak structure disappears and a single peak of
large magnitude develops instead. Overall, these conclusions are
similar to the case of shearing and can be summarized as follows.
Tensile loading induces deformation and coalescence of compact pores
into larger voids that ultimately lead to formation of
system-spanning empty regions associated with breaking of samples in
two parts. The process of large pores formation is consistent with
series of system snapshots shown in
Figs.\,\ref{fig:snapshot_strain_rho03}--\ref{fig:snapshot_strain_rho08_slice},
where the material's failure is accompanied with pore redistribution
into larger domains and with densification of the solid parts.

\vskip 0.05in


The evolution of the pore-size distributions with applied strain
during mechanical loading is important for overall understanding
response properties of porous materials.  However, as any average
quantity, they do not provide any spatially-resolved information on
the dynamic events in material systems under loading. In what
follows, we therefore augment the PSDs by spatially-resolved density
profiles.  Specifically, we consider locally averaged density,
computed along the direction of mechanical loading, $\langle \rho
\rangle_{s}(x)$. This quantity is defined as the number of atoms
located in a thin slice of thickness $\sigma$ along the $\hat{x}$
direction (i.e., the direction of mechanical loading), divided by
the volume of the slice: $L_{y}L_{z}\sigma^3$ (where $L_y$ and $L_z$
are the system sizes in the two Cartesian directions perpendicular
to the loading direction).

\vskip 0.05in

\begin{figure}[t]
\includegraphics[width=12.cm,angle=0]{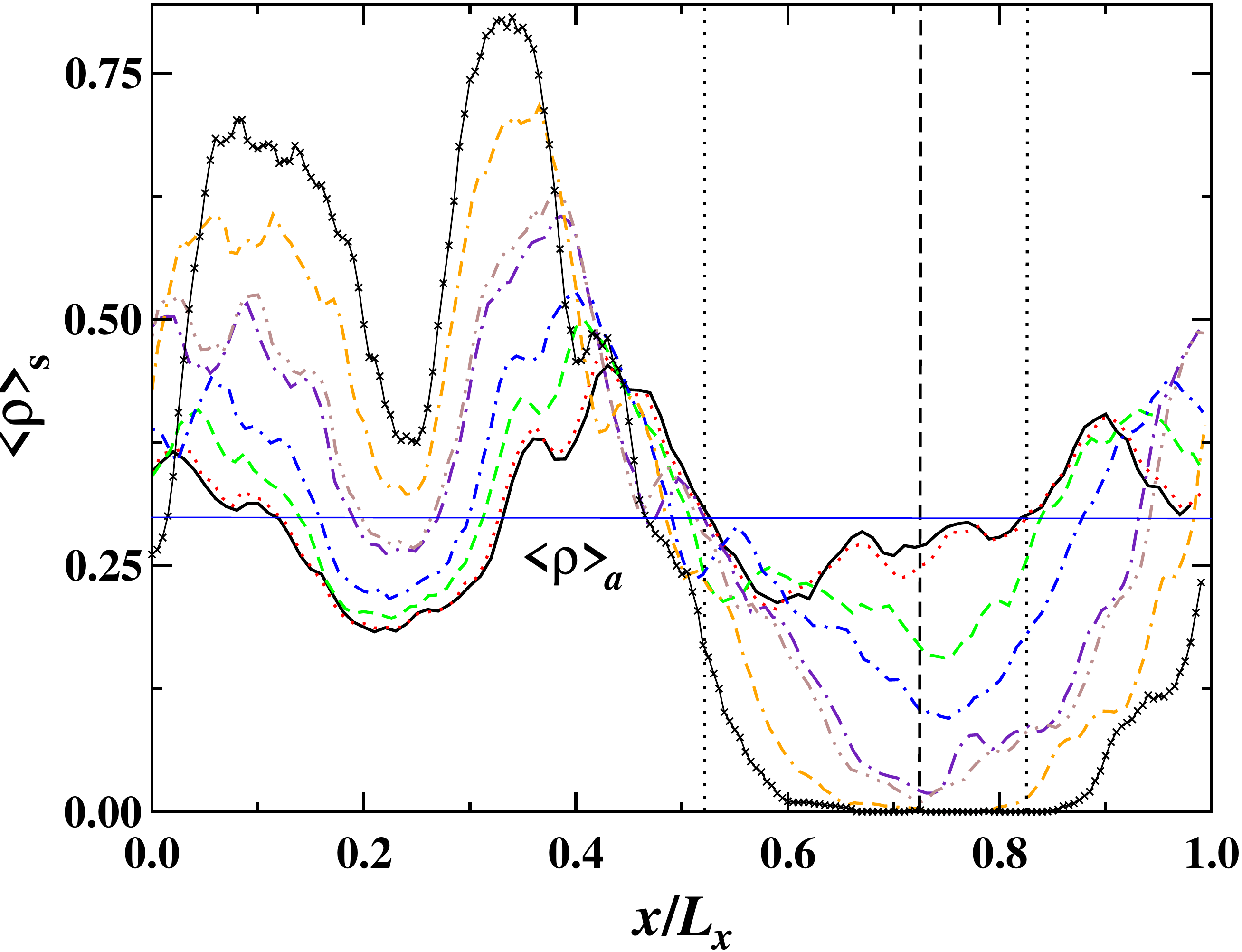}
\caption{(Color online) The density profiles $\langle \rho
\rangle_{s}(x)$ (in units of $\sigma^{-3}$) for the values of strain
$\varepsilon_{xx}=0.0$ (solid black curve), $0.05$ (dotted red
curve), $0.15$ (dashed green curve), $0.25$ (dash-dotted blue
curve), $0.45$ (dash-dotted violet curve), $0.5$ (dash-double-dotted
brown curve), $0.75$ (dash-dotted orange curve), and $1.0$ (black
curve with crosses).  The data were averaged in one sample in thin
slices parallel to the $yz$ plane.   The horizontal blue line
indicates the average glass density $\rho\sigma^{3}=0.3$.  The two
vertical dotted lines mark the borders of the region with reduced
density.  The dashed vertical line shows the position of the failure
zone center.}
\label{fig:den_prof_rho03}
\end{figure}
\begin{figure}[t]
\includegraphics[width=12.cm,angle=0]{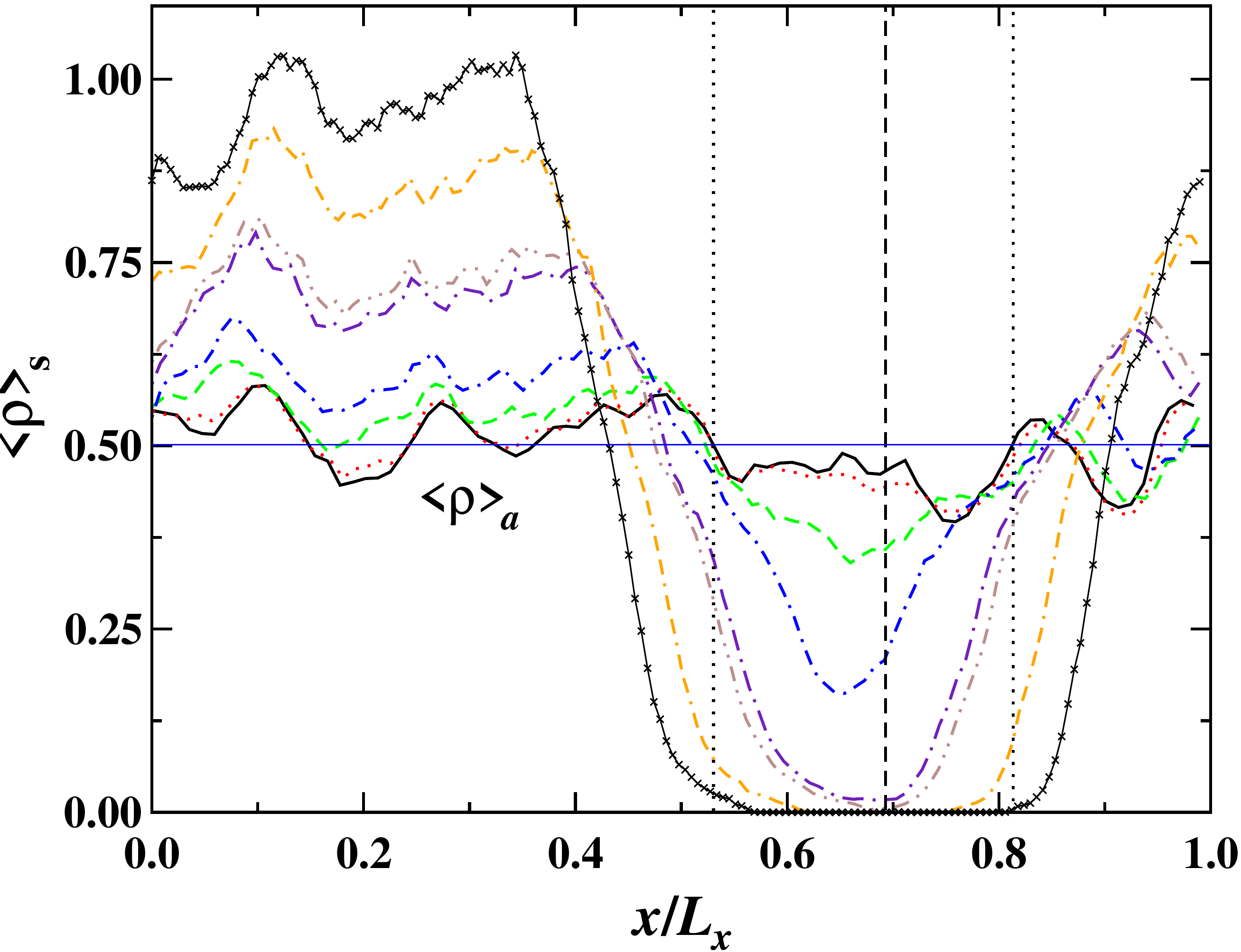}
\caption{(Color online) The atomic density profiles $\langle \rho
\rangle_{s}$ (in units of $\sigma^{-3}$) along the $\hat{x}$ axis
for the same values of strain as in Fig.\,\ref{fig:den_prof_rho03}.
The average glass density $\rho\sigma^{3}=0.5$ is indicated by the
horizontal blue line. The colorcode is the same as in
Fig.\,\ref{fig:den_prof_rho03}.  }
\label{fig:den_prof_rho05}
\end{figure}
\begin{figure}[t]
\includegraphics[width=12.cm,angle=0]{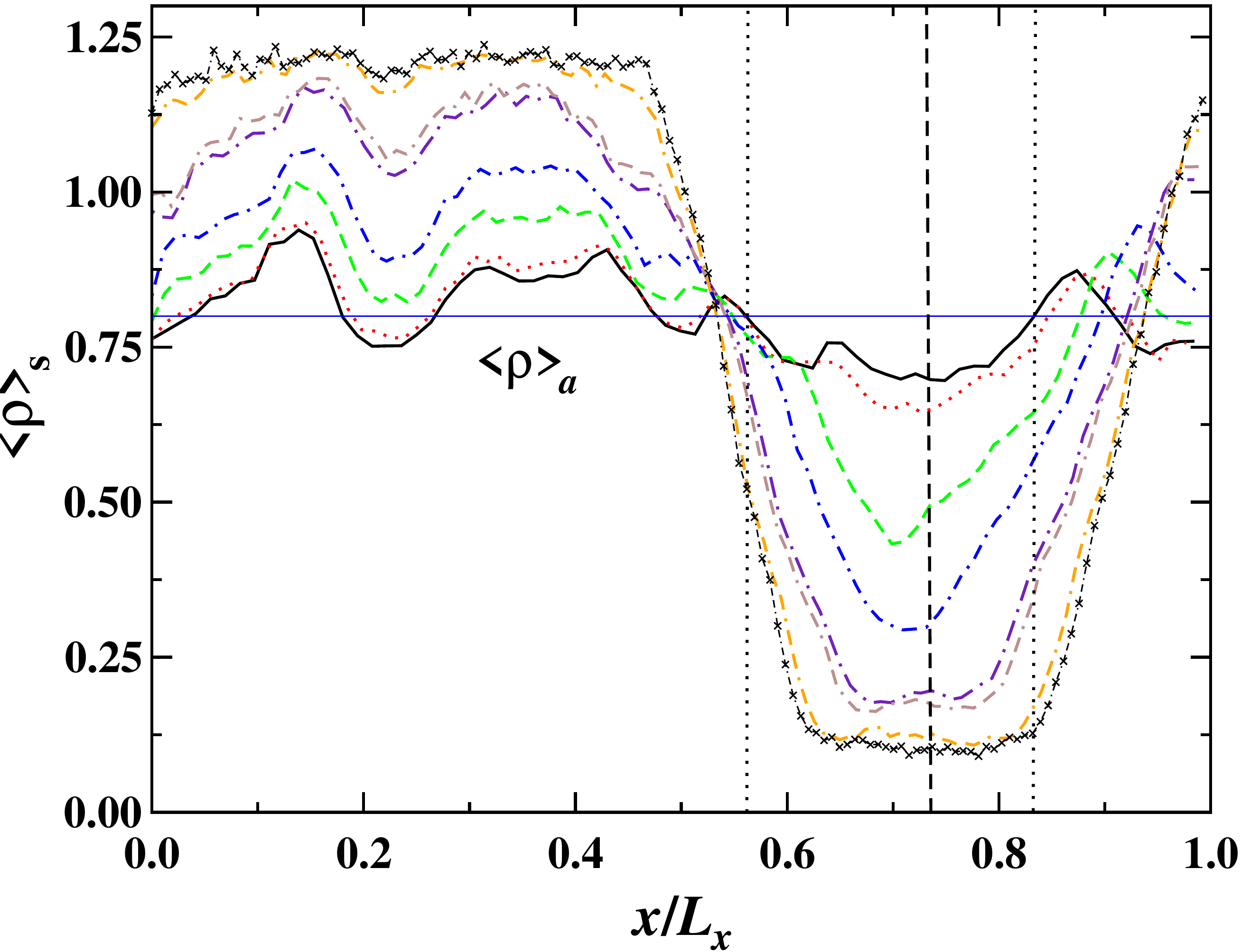}
\caption{(Color online) The averaged density profiles $\langle \rho
\rangle_{s}(x)$ (in units of $\sigma^{-3}$) for selected values of
strain. The average glass density is $\rho\sigma^{3}=0.8$. The
colorcode and strain values are the same as in
Figs.\,\ref{fig:den_prof_rho03} and \ref{fig:den_prof_rho05}.}
\label{fig:den_prof_rho08}
\end{figure}

The temporal evolution of the average density profiles in porous
samples is illustrated in Figs.\,\ref{fig:den_prof_rho03},
\ref{fig:den_prof_rho05} and \ref{fig:den_prof_rho08}. Here, we
present the results for systems with reduced densities,
$\rho\sigma^3$, of $0.3$, $0.5$, and $0.8$. First and foremost, we
would like to emphasize one common feature, which is characteristic
of all the samples, we studied in this work.  It regards the
location of the zone, where material's failure occurs.  As follows
from Figs.\,\ref{fig:den_prof_rho03}--\ref{fig:den_prof_rho08}, the
location of the zone is correlated with the region, where density is
lower than its average over the whole sample. However, a deeper
analysis suggests that the defining factor is \textit{the spatial
extent of such a region}, rather then the absolute value of local
density deviation from the average.  Moreover, the center of the
low-density region approximately coincides with the location of the
failure. In other words, a local deviation of density from its
average in a narrow spatial region does not signifies a weak (from
the mechanic's perspective) region. Rather, the failure takes place
in the center of extended low-density zone.

\vskip 0.05in


This is evident in all three cases of the average glass density
presented in Figs.\,\ref{fig:den_prof_rho03},
\ref{fig:den_prof_rho05} and \ref{fig:den_prof_rho08}. Indeed, at
strains around $0.1$, a dip in $\langle \rho \rangle_{s}(x)$ starts
to develop within the regions with low average densities.   The
process of local density decrease in these regions is accompanied by
simultaneous densification in the neighboring parts of the systems.
Note that shapes of the density patterns are largely preserved in
the remote - from the regions with large paucity - parts of the
systems. In loose terms, one can say that the patterns repeat
themselves, the only difference being their magnitudes and lateral
shift due to gradual increase in the extent of the low-density
region.  Within each low-density region, in the initial stages of
loading, the density profile shows a rather sharp dip. In the later
stages, the profiles show the characteristic for interfacial regions
(hyperbolic-tangent like) shapes.

\vskip 0.05in


We finally comment that the process of density evolution is gradual;
i.e., no abrupt transitions between density states were observed.
The cases of strains $\varepsilon_{xx}=0$ and $0.05$ as well as
$\varepsilon_{xx}=0.45$ and $0.50$ in
Figs.\,\ref{fig:den_prof_rho03}--\ref{fig:den_prof_rho08} provide an
illustration of the premise in the initial and intermediate stages
of loading, correspondingly. Upon failure, the samples with high
densities exhibit some density relaxation on both sides of the
failure region. Furthermore, a part of the tensile energy, supplied
in the process of external loading, is stored in the dense parts of
the system. This follows from the temporal behavior of the systems
after failure. The two parts continue to densify after break-up, as
evident from the behavior of systems with $\rho\sigma^3 = 0.3$ and
0.5 shown in Figs.\,\ref{fig:den_prof_rho03} and
\ref{fig:den_prof_rho05}, respectively. In the case of system with
high densities (see $\rho\sigma^3 = 0.8$) the effects is less
pronounced. Note, however, the behavior in the region around density
dip located at $x/L_x \sim 0.22$ in Fig.\,\ref{fig:den_prof_rho08}.
There is clear sign of continued relaxation due to elastic energy,
accumulated during the loading. The magnitude of the effects is
expected to be smaller, as structural rearrangement in a
high-density material requires more energy, as compared to its
low-density counterparts.

\section{Conclusions}

In summary, the mechanical response and transformation of porous
structure in binary glasses under tension were studied using
molecular dynamics simulations. The binodal glassy systems were
prepared via rapid quench from a liquid state to a temperature well
below the glass transition. In the process of phase separation at
constant volume, different pore morphologies are formed depending on
the average glass density. Visual observation of system snapshots in
the absence of deformation shows atom configurations with randomly
distributed, isolated pores at higher average glass densities, while
interconnected porous structure is formed at lower densities.  These
structural changes are reflected in the shape of pore size
distribution functions computed at different average glass
densities.

\vskip 0.05in

When the porous material is subjected to tensile loading at constant
volume, the stress-strain curves exhibit a linear regime, where the
absolute value of stress increases up to the yield point, and then
followed by plastic deformation at large strain until failure.
Consistently with theoretical predictions, the power-law exponent of
the elastic modulus dependence on the average glass density is the
same as for the shear modulus of porous glasses reported in our
previous study~\cite{Priezjev17}. Furthermore, under tensile
loading, the pores become significantly deformed and redistributed
spatially, thus forming a system-spanning void associated with
breaking of the amorphous material. The analysis of the locally
averaged density profiles elucidates the mechanism of the failure
mode which originates in the extended regions of lower glass
density.

\section*{Acknowledgments}

Financial support from the National Science Foundation (CNS-1531923)
is gratefully acknowledged. The study has been in part funded by the
Russian Academic Excellence Project `5-100'. The molecular dynamics
simulations were performed using the LAMMPS numerical code developed
at Sandia National Laboratories~\cite{Lammps}.  Computational work
in support of this research was performed at Michigan State
University's High Performance Computing Facility and the Ohio
Supercomputer Center.


%

%

%

%

%

%

%

%

%

%

%

%

%

\bibliographystyle{prsty}

\end{document}